\documentclass{article}

\usepackage{arxiv}

\usepackage[utf8]{inputenc} 
\usepackage[T1]{fontenc}    
\usepackage{hyperref}       
\usepackage{url}            
\usepackage{booktabs}       
\usepackage{amsfonts}       
\usepackage{nicefrac}       
\usepackage{microtype}      
\usepackage{lipsum}

\usepackage{amsfonts}
\usepackage{amssymb}
\usepackage{bbold}
\usepackage{booktabs}
\usepackage{ulem}
\usepackage{xr}
\usepackage[justification=centering ]{subcaption}
\usepackage[numbers]{natbib} 
\usepackage{graphicx}
\usepackage{amsmath}

\newcommand{\effi}{f_i}
\renewcommand{\vec}[1]{\ensuremath{\mathbf{#1}}}
\newcommand{\Ca}{\text{Ca}}
\newcommand{\CaTB}{\mbox{Ca}^{\tiny\mbox{TB}}}
\newcommand{\CaTT}{\mbox{Ca}^{\tiny\mbox{TT}}}

\newcommand{\muin}{\mu_{\mbox{\scriptsize in}}}
\newcommand{\muout}{\mu_{\mbox{\scriptsize out}}}
\newcommand{\mum}{\mu_{\mbox{\scriptsize m}}}

\newcommand{\Kal}{k_{\scriptsize \alpha}}

\newcommand{\kB}{k_{\mbox{\scriptsize B}}}
\newcommand{\kS}{k_{\mbox{\scriptsize S}}}

\newcommand{\We}{W_{\mbox{\scriptsize e}}}
\newcommand{\ws}{w_{\mbox{\scriptsize S}}}

\newcommand{\WB}{W_{\mbox{\scriptsize B}}}
\newcommand{\WS}{W_{\mbox{\scriptsize S}}}

\newcommand{\mus}{\mu_{\mbox{\scriptsize s}}}
\newcommand{\mud}{\mu_{\mbox{\scriptsize d}}}

\newcommand{\gammadot}{\dot{\gamma}}

\usepackage[labelfont=bf,textfont=normalfont,singlelinecheck=off,justification=raggedright]{subcaption}

\title{Lattice Boltzmann simulations on the tumbling to tank-treading transition: effects of membrane viscosity}

\author{
  Fabio Guglietta\\ 
  Department of Physics \& INFN, {\it University of Rome ``Tor Vergata''}\thanks{Via della Ricerca Scientifica 1, 00133, Rome, Italy}\\ 
Chair for Computational Analysis of Technical Systems (CATS), {\it RWTH Aachen University}\thanks{52056 Aachen, Germany}\\
Computation-based  Science  and  Technology  Research  Center, {\it The  Cyprus  Institute}\thanks{20 Konstantinou Kavafi Str., 2121 Nicosia, Cyprus}\\
  \texttt{fabio.guglietta@roma2.infn.it} \\
   \And
 Marek Behr\\
 Chair for Computational Analysis of Technical Systems (CATS), {\it RWTH Aachen University}$^{\dagger}$\\
  \And
 Luca Biferale\\
  Department of Physics \& INFN, {\it University of Rome ``Tor Vergata''}$^{*}$\\ 
  \And
 Giacomo Falcucci\\
 Department of Enterprise Engineering “Mario Lucertini,” {\it University of Rome ``Tor Vergata"}\thanks{Via del Politecnico 1, 00133, Rome, Italy}\\ 
 \AND
 Mauro Sbragaglia\\
  Department of Physics \& INFN, {\it University of Rome ``Tor Vergata''}$^{*}$\\ 
}

\begin{document}
\maketitle
 \vskip 0.4in
\begin{abstract}
The tumbling to tank-treading (TB-TT) transition for red blood cells (RBCs) has been widely investigated, with a main focus on the effects of the viscosity ratio $\lambda$ (i.e., the ratio between the viscosities of the fluids inside and outside the membrane) and the shear rate $\gammadot$ applied to the RBC. However, the membrane viscosity 
$\mum$ plays a major role in a realistic description of RBC's dynamics, and only a few works have systematically focused on its effects on the TB-TT transition. In this work, we provide a parametric investigation on the effect of membrane viscosity $\mum$ on the TB-TT transition, for a single RBC. 
It is found that, at fixed viscosity ratios $\lambda$, larger values of $\mum$ lead to an increased range of values of capillary number at which the TB-TT transition occurs. 
We systematically quantify such an increase by means of mesoscale numerical simulations based on the lattice Boltzmann models.
\end{abstract}

\section{Introduction}
Red blood cells (RBCs) are highly deformable cells that are immersed in a Newtonian fluid called plasma, and they constitute the most important part of blood: their concentration, called the hematocrit, ranges between 37\% and 50\%~\cite{thesis:kruger}. Thus, blood can be considered as a dense suspension of highly deformable particles (RBCs) in plasma and  their dynamics is crucial to dissect blood flow phenomena. 
Due to their deformability, RBCs give rise to    different dynamics as compared to that related to rigid particles.
In 1972, Goldsmith \& Marlow~\cite{goldsmith1972flow} devised an experiment to study the dynamics of RBCs at low shear rates. In this regime, they found that erythrocytes ``tumble'', with a rigid body-like behaviour: for this reason, this dynamics is called {\it tumbling motion} (TB). 
On the other hand, at high shear rates, the RBC membrane rotates while the cell keeps a fixed inclination with respect to the flow direction, providing the so-called {\it tank-treading motion} (TT)~\cite{fischer1978red}. 
Decreasing the value of the shear rate leads to a reduction in the cell deformation, as well as in its inclination with respect to the flow: in this configuration, the membrane rotates like in the TT motion, while the cell inclination fluctuate in time~\cite{abkarian2007swinging}. This dynamics is called {\it swinging motion}. 
Apart from these regimes, RBCs show several other different dynamics, like the rolling, frisbee motion, trilobe dynamics, etc. (see~\cite{viallat2019dynamics} for a review), depending on several parameters (the capillary number, the viscosity ratio, the orientation with respect to the shear plane, among the others). 
In the present work, we focus on the tumbling (TB) and tank-treading (TT) motions.  \\
A basic model to understand the tumbling-to-tank-treading (TB-TT) transition has been developed by Keller \& Skalak~\cite{keller1982motion} (KS-model); it considers the motion of a pure viscous ellipsoid in simple shear flow, assuming that the particle does not change shape during its motion. 
This model addresses the dependency of the TB-TT transition on the viscosity ratio 
$\lambda$, while it is not capable of predicting two behaviours experimentally observed by Abkarian {\it et al.}~\cite{abkarian2007swinging}: the dependency of the TB-TT transition on the shear rate $\gammadot$, and the swinging motion. To explain the dependency on $\gammadot$, Abkarian {\it et al.}~\cite{abkarian2007swinging} and Skotheim \& Secomb~\cite{skotheim2007red} introduced an elastic energy contribution in the membrane model: in the presence of shear rates $\gammadot$, the flow energy is in part dissipated by viscous friction inside the cell and on the membrane (as in KS-model) and in part stored as elastic energy.
When $\gammadot$ is below a critical value $\gammadot_c$, the injected energy is insufficient to trigger tank-treading motion~\cite{skotheim2007red}, and the cell behaves like a rigid body (i.e., it \textit{tumbles}). For $\gammadot \geq \gammadot_c$, the membrane starts to tank-tread and the inclination angle oscillates.\\ 
Since energy dissipation assumes a key role in the TB-TT transition, it is crucial to account for the membrane viscosity in order to give a realistic description of the erythrocyte dynamics. Even though several works explore the RBC TB-TT transition as related to the shear rate $\gammadot$ and the viscosity ratio $\lambda$ (see ~\cite{skotheim2007red,abkarian2007swinging,kruger2013crossover,dupire2012full} and references therein), only few works focus on the effect of membrane viscosity $\mum$. This paper aims at taking some steps further to fill this gap. In our previous work~\cite{D0SM00587H}, the effect of membrane viscosity during the relaxation after the cessation of a mechanical load was investigated; here, the quantitative effect of the membrane viscosity on the TB-TT transition is explored, adopting the same numerical model as in~\cite{D0SM00587H}. More specifically, the Skalak model~\cite{art:skalaketal73} and the Helfrich formulation~\cite{art:helfrich73} are implemented to describe the elastic behaviour, while the Standard Linear Solid model is employed to account for the membrane viscosity~\cite{art:lizhang19}; the fluid is resolved in the framework of the Lattice Boltzmann Method (LBM)~\cite{thesis:kruger}.

\section{Model description}\label{sec:model}
\begin{figure}
\centering
 \begin{subfigure}[t]{.8\linewidth}
 \includegraphics[width=1.\linewidth]{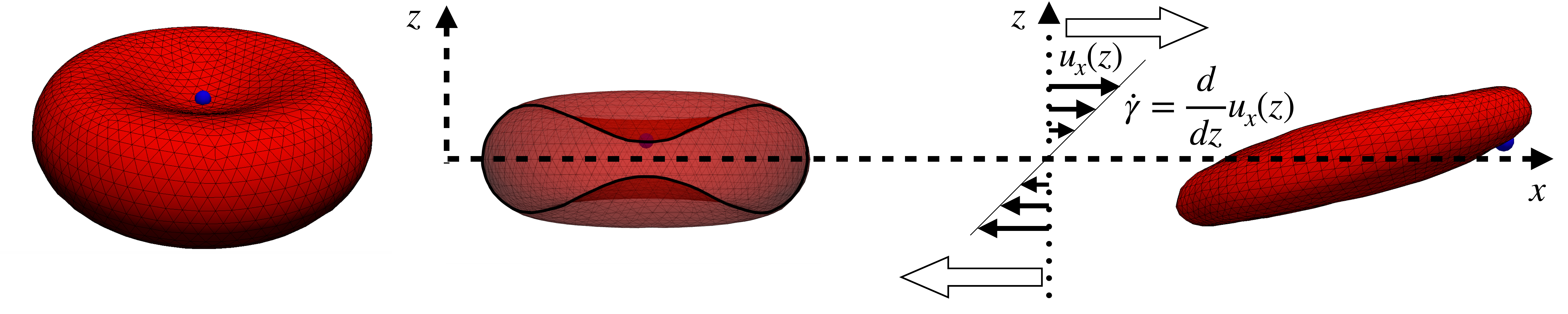}\caption{{\it Left}: 3D triangular mesh representing the RBC. 
{\it Right}: RBC at rest and in shear flow with intensity $\gammadot$. In both panels, the little blue sphere is marked to highlight the membrane element chosen to identify the angle $\phi$. \label{fig:sketch-a}}
\end{subfigure}
 \begin{subfigure}[t]{.8\linewidth}
 \includegraphics[width=1.\linewidth]{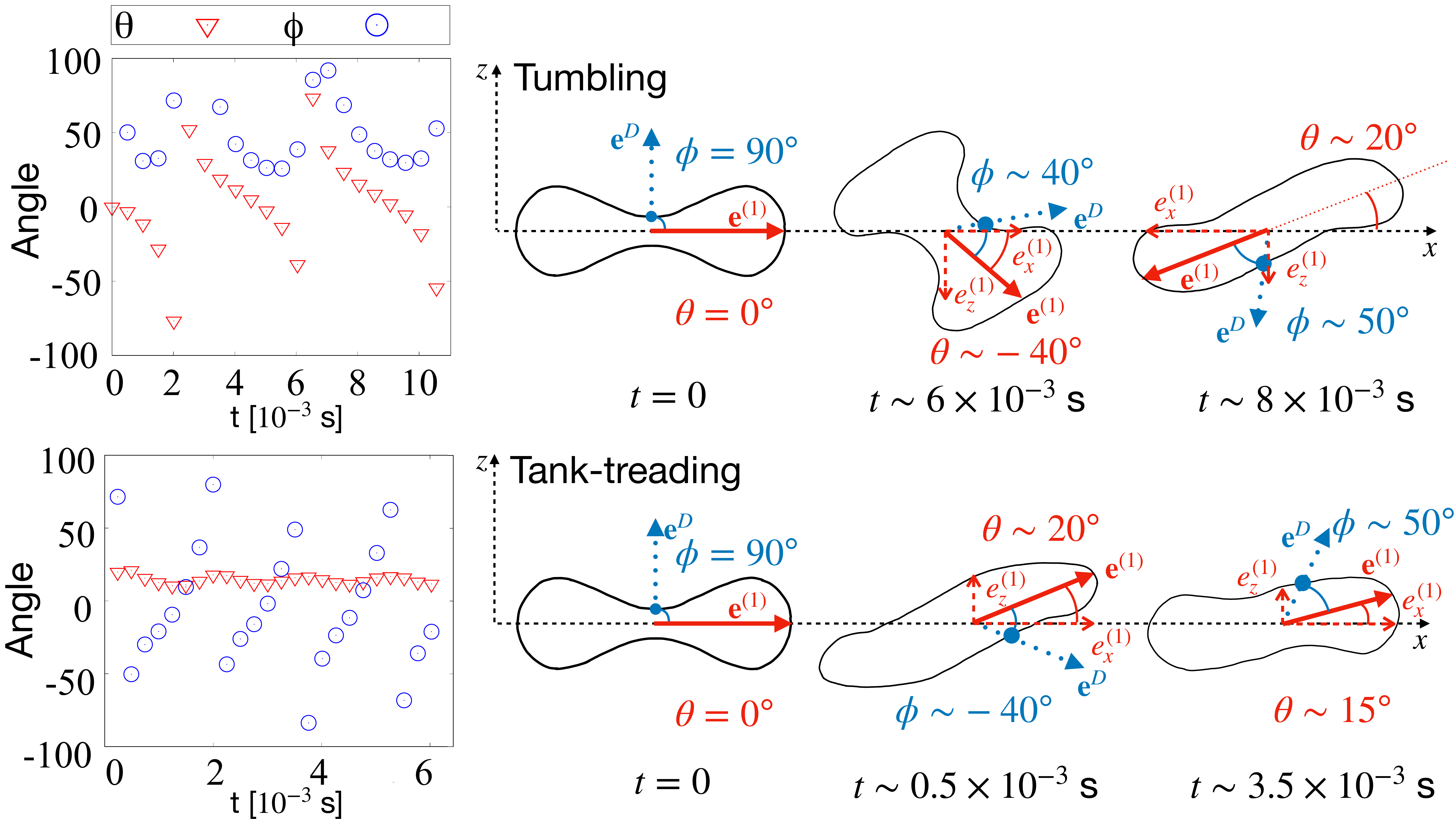}\caption{We report two examples of tumbling (TB) and tank-treading (TT): on the left, the plots with the time evolution of $\theta$ and $\phi$ are reported; on the right, we report three significant snapshots taken at different times $t$ (values of the angles $\theta$ and $\phi$ are also reported). The two vectors $\vec{e}^{(1)}$ (continuous-line arrow) and $\vec{e}^D$ (dotted-line arrow) used to identify the angles $\theta$ and $\phi$, respectively, are also displayed.} \label{fig:sketch-b}
\end{subfigure}
\caption{Sketch of the 3D numerical simulations performed. \label{fig:sketch}}
\end{figure}

\begin{figure}[t!]
\centering
 \includegraphics[width=.5\linewidth]{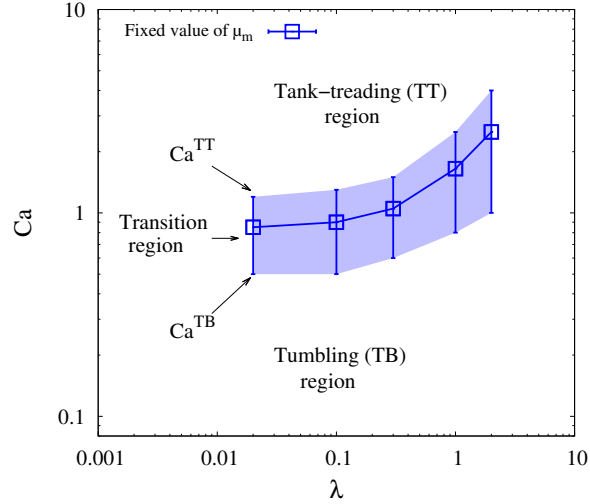}
 \caption{Sketch of phase diagram for the shear plane dynamics. The TB region and the TT region (below and above the shaded regions, respectively) correspond to the values of capillary number $\Ca$ and viscosity ratio $\lambda$ where we found a pure TB and TT, respectively.  
 The transition region (shaded) is identified by the highest ($\CaTB$) and lowest ($\CaTT$) value of $\Ca$ at which a pure TB and TT are found, respectively. The error bars represent the width of the transition region $\Delta\Ca$ (see Eq.~\eqref{eq:deltaca}). In Fig.~\ref{fig:trans-a} and Fig.~\ref{fig:trans2}, since we report results for different values of membrane viscosity $\mum$, to avoid the superposition of shaded regions we represent them only by error bars. Data reported here are dummy, and they are used to sketch the plots showed in Fig.~\ref{fig:trans-a} and Fig.~\ref{fig:trans2}.}
 \label{fig:sketch_trans}
\end{figure}

\begin{figure}[t!]
\centering
\begin{subfigure}[t]{.45\linewidth}
 \includegraphics[width=1.\linewidth]{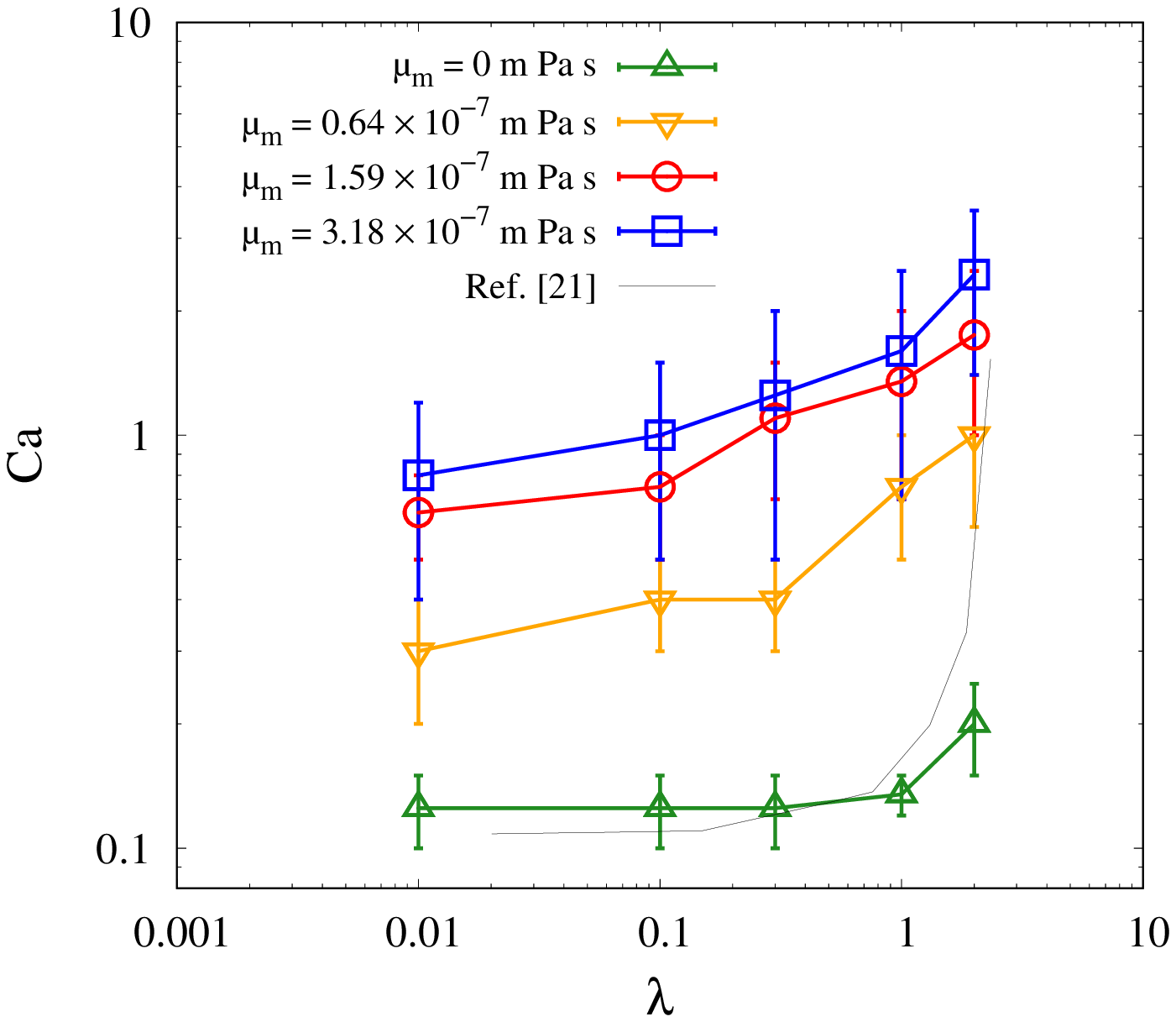}\caption{\label{fig:trans1-a}}
\end{subfigure}
\begin{subfigure}[t]{.45\linewidth}	
 \includegraphics[width=1.\linewidth]{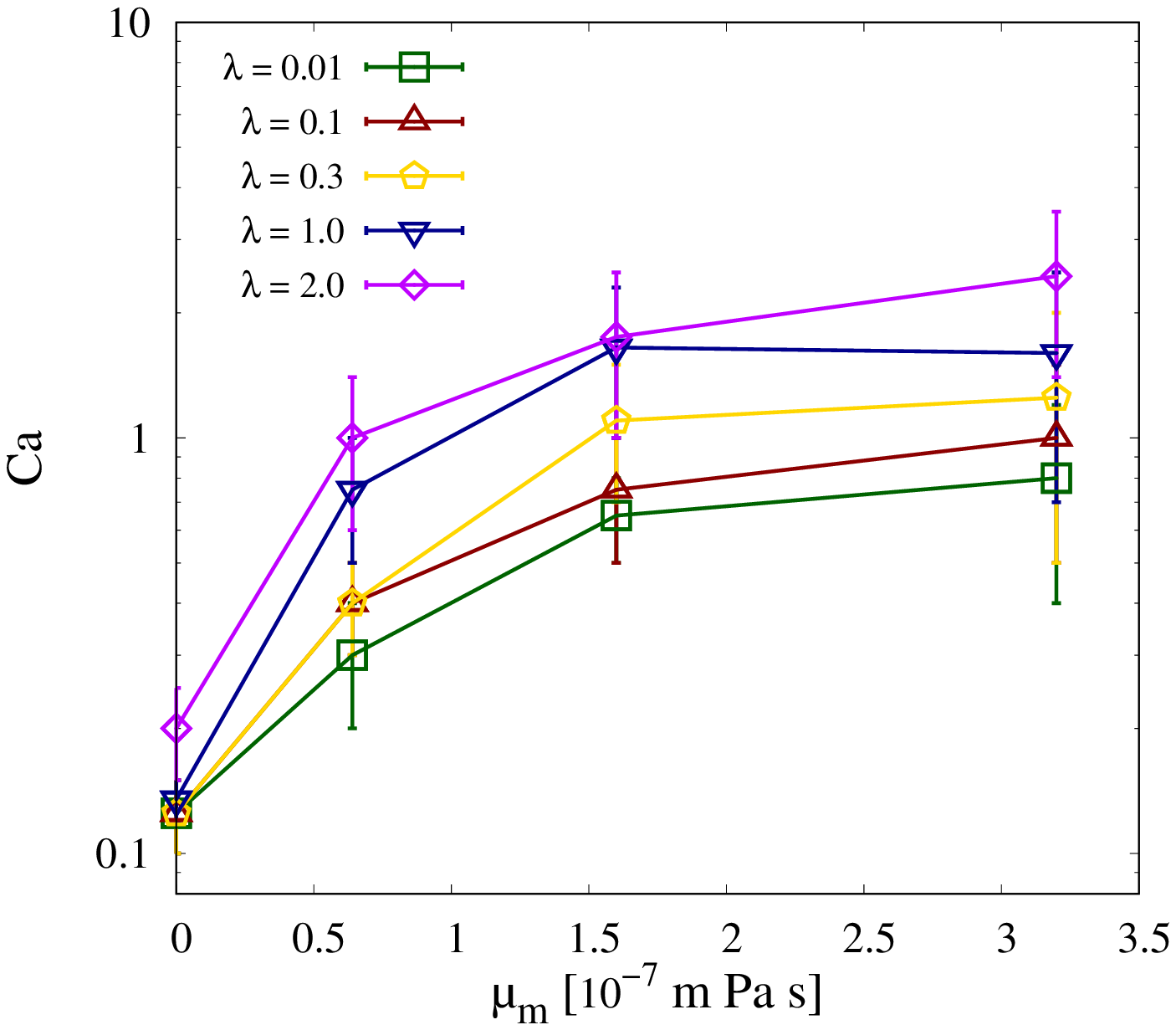}\caption{\label{fig:trans1-b}}
\end{subfigure}
 \caption{Phase diagram for the shear plane dynamics. {\it Panel a}: we report results for four different values of membrane viscosity $\mum$, from $\mum=0$ to $\mum = 3.18\times 10^{-7}\mbox{ m Pa s}$, at varying viscosity ratio $\lambda$. Error bars represent the width of the transition region $\Delta\Ca$ (see Eq.~\eqref{eq:deltaca}, and Fig.~\ref{fig:sketch_trans}). Numerical results from Cordasco {\it et al.}~\cite{cordasco2014comparison} are also reported for $\mum=0$ (the black line with no points is the transition line from TB to TT). {\it Panel b}: we report results for five different values of viscosity ratio $\lambda$, form $\lambda=0.01$ to $\lambda = 2$, at varying viscosity ratio $\mum$. \label{fig:trans-a}}
\end{figure}

\begin{figure}[t!]
\centering
\begin{subfigure}[t]{.45\linewidth}
 \includegraphics[width=1.\linewidth]{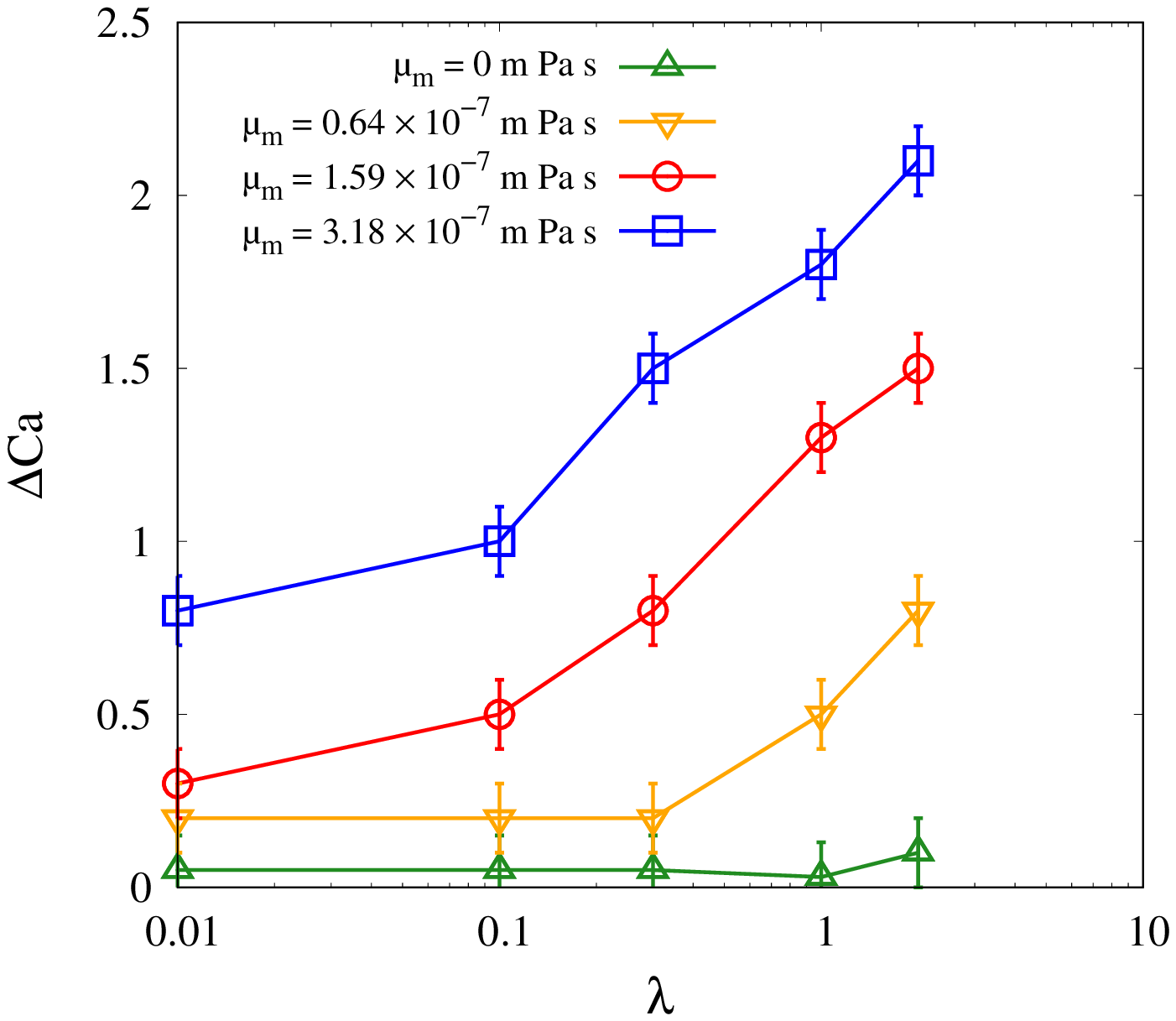}\caption{\label{fig:trans2-a}}
\end{subfigure}
\begin{subfigure}[t]{.45\linewidth}	
 \includegraphics[width=1.\linewidth]{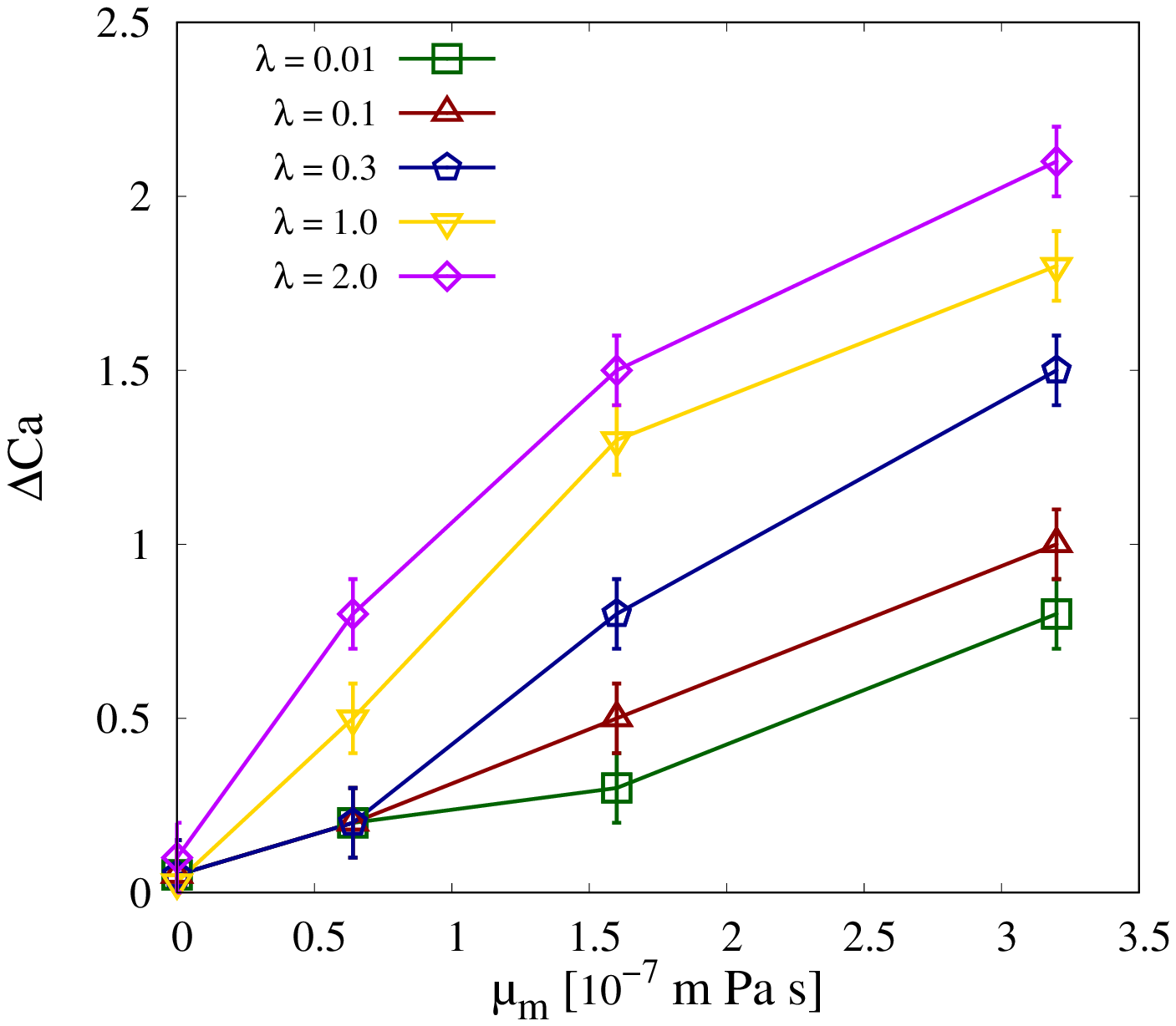}\caption{\label{fig:trans2-b}}
\end{subfigure}
\caption[figure]{We report the width of the transition region $\Delta\Ca$ (see Eq.~\eqref{eq:deltaca}, and Fig.~\ref{fig:sketch_trans}) as a function of viscosity ratio $\lambda$ ({\it panel~a}) and as a function of membrane viscosity $\mum$ ({\it panel~b}).\label{fig:trans2}}
\end{figure}

In this work, the RBC membrane is represented by a 3D triangular mesh, and its shape at rest is the typical biconcave shape described by Evans \& Fung~\cite{evans1972improved}:
\begin{equation}
z(x,y) = \pm \sqrt{1-\frac{x^2+y^2}{r^2}}\left(C_0+C_1\frac{x^2+y^2}{r^2}+C_2\left(\frac{x^2+y^2}{r^2}\right)^2\right)\; ,
\end{equation}
with $C_0 = 0.81\times 10^{-6} \mbox{ m}$, $C_1 = 7.83\times 10^{-6} \mbox{ m}$ and $C_2 = -4.39\times 10^{-6} \mbox{ m}$; $r=3.91\times 10^{-6} \mbox{ m}$ is the large radius (details on the conversion between physical and lattice units are provided in~\cite{thesis:kruger,book:kruger}).\\
Since the membrane has a thickness of about $4\times 10^{-9}$ m~\cite{book:gommperschick}, the RBC is considered as a 2D viscoelastic membrane filled with a Newtonian fluid: we adopt the {\it Skalak model} to describe the resistance to shear and area deformations~\cite{thesis:kruger}, and the {\it Helfrich formulation} for the bending resistance~\cite{thesis:kruger}. The viscous behaviour is described by the {\it Standard Linear Solid} (SLS) model, in which every element of the mesh can be thought of as characterised by a dashpot and an artificial spring connected in series that are connected together in parallel with another spring~\cite{art:lizhang19}. For each node of the mesh, we compute the total force that is the sum of each of the above viscoelastic contributions.\\
In detail, for the elastic contribution, we compute the free energy of the membrane $\We = \WS+\WB$ that corresponds to the strain and bending energy, respectively. According to Skalak model, we have
\begin{equation}
\WS = \sum_j \ws^{(j)}A_j\; ,
\end{equation}
where $A_j$ is the area of the $j-$th element of the triangular mesh, and $\ws^{(j)}$ is the area energy density related to the $j$-th element, given by:
\begin{equation}\label{eq:skalak2}
\ws= \frac{\kS}{8}\left(I_1^2+2I_1-2I_2\right) +  \frac{\Kal}{8} I_2^2\; ,
\end{equation}
where $I_1 = \lambda_1^2+\lambda_2^2-2$ and $I_2 = \lambda_1^2\lambda_2^2-1$ are the strain invariants for the $j$-th element, while $ \lambda_1$ and $ \lambda_2$ are the principal stretch ratios~\cite{art:skalaketal73,thesis:kruger}. Eq.~\eqref{eq:skalak2} is made of two terms: the first one describes shear deformations, and the related elastic modulus is $\kS$; the second one is introduced to describe area dilation, and the corresponding elastic modulus is $\Kal$. Note that the surface elastic shear modulus $\kS$ enters in the Capillary number $\Ca$ defined as:
\begin{equation}\label{eq:ca}
\Ca = \frac{\muout\dot{\gamma}r}{\kS}\; .
\end{equation}
The bending energy can be discretised as follow~\cite{thesis:kruger}:
\begin{equation}\label{eq:helfrich}
\WB = \frac{\kB\sqrt{3}}{2}\sum_{\langle i,j\rangle}\left(\theta_{ij}-\theta_{ij}^{(0)}\right)^2\; ,
\end{equation}
where $\theta_{ij}$ is the angle formed by the normals of the $i$-th and $j$-th faces of the triangular mesh (the superscript (0) refers to the angle at rest); the sum runs over the neighbouring faces $i,j$; $\kB$ is the bending modulus. Once we have computed the elastic free energy $\We$ for all faces, the force on the node $i$-th can be computed as the derivative of the free energy $\We$ with respect to the position of the node $\vec{x}_i$: 
\begin{equation}\label{eq:nodal_force_energy}
\vec{F}_i = -\frac{\partial \We(\vec{x}_i)}{\partial \vec{x}_i}\; , 
\end{equation}
where $\We(\vec{x}_i)$ is the sum of the free energy of all faces sharing the node $i$-th.\\
Regarding the viscous part, we first compute the 2D viscous stress given by
\begin{equation}\label{eq:mv2}
\pmb{\tau}^\nu = \mus \left(2\dot{\vec{E}} -\mbox{tr}(\dot{\vec{E}})\mathbb{1}\right) + \mud \mbox{tr}(\dot{\vec{E}})\mathbb{1}\; ,
\end{equation}
where $\dot{\vec{E}}$ is the strain rate tensor; $\mus$ and $\mud$ are shear and dilatational viscosities, respectively. In this work, we consider $\mus=\mud=\mum$~\cite{barthes1985}. Note that $\mum$ is the viscosity of the 2D membrane, and then it is measured in [m Pa s]. After having computed the viscous stress tensor $\pmb{\tau}^\nu$, the force acting on the $i$-th node can be computed as
\begin{equation}\label{eq:nodal_force}
\vec{F}_i=\pmb{\tau}^\nu \pmb{\mathcal{F}}^{-T}\pmb{\nabla} N_i A_m\; ,
\end{equation}
where $\pmb{\mathcal{F}}$ is the gradient deformation tensor which is contracted with the viscous tensor $\pmb{\tau}^\nu$, $\pmb{\nabla} N_i$ is the gradient of the shape functions and $A_m$ is the surface area of the $m$-th face of the triangular mesh~\cite{thesis:kruger,gounley2015computational}.
More details are given in~\cite{art:lizhang19,D0SM00587H}.\\
In order to simulate the dynamics of a single RBC in simple shear flow, we use the {\it Lattice Boltzmann Method} (LBM) to solve the fluid, and the {\it Immersed Boundary Method} (IBM) to describe the interaction between the fluid and the membrane~\cite{book:kruger}. 
The LBM hinges on the Lattice Boltzmann equation~\cite{book:kruger}:
\begin{equation}\label{LBMEQ}
\effi(\vec{x}+\vec{c}_i\Delta t, t+ \Delta t) - \effi(\vec{x}, t) = -\frac{\Delta t}{\tau}\left(\effi(\vec{x}, t) - \effi^{(\mbox{\tiny eq})}(\vec{x}, t)\right) + \effi^{(F)}\; ,
\end{equation}
where $\effi(\vec{x}, t)$ is the probability density function of fluid molecules with discrete velocity $\vec{c}_i$ at position $\vec{x}$ and at time $t$, $\Delta t$ is the discrete time interval, $\tau$ is the relaxation time, $\effi^{(\mbox{\tiny eq})}$ is the equilibrium distribution function (that is the analogous of the Maxwell distribution in the Boltzmann equation), and $\effi^{(F)}$ is the force density that has been implemented according to the Guo scheme~\cite{PhysRevE.65.046308}. Once the 	probability density function $\effi(\vec{x}, t)$ has been computed, one can compute the density and the velocity fields: $\rho(\vec{x}, t) = \sum_{i} \effi(\vec{x}, t)$ and $\vec{u}(\vec{x}, t) = \sum_{i} \vec{c}_i \effi(\vec{x}, t)/\rho(\vec{x}, t)$, respectively~\cite{book:kruger}.\\
We consider two different viscosities for the two fluids inside and outside the membrane, whose ratio is given by 
\begin{equation}\label{eq:visc_ratio}
\lambda = \frac{\muin}{\muout} \; .
\end{equation}
In order to distinguish which lattice sites lie inside or outside the membrane, we have implemented the (parallel) Hoshen-Kopelmann algorithm~\cite{art:frijters15}.\\
The IBM first couples the membrane to the fluid via an interpolation of the nodal forces of the 3D triangular mesh on the lattice sites; then, interpolating the fluid velocity on the mesh nodes we get the coupling between the fluid and membrane~\cite{book:kruger}. The interpolation is performed by means of discrete Delta functions (in particular, we use a 4-point stencil). For more details on the model adopted, we refer to our previous work~\cite{D0SM00587H}.\\
Finally, we compute the angles $\theta$ and $\phi$ (see Fig.~\ref{fig:sketch}) in the following way: $\theta$ is the angle that the longest eigenvector $\vec{e}^{(1)}$ of the inertia tensor forms with the $x-$axis, and it is positive if $e_x^{(1)}e_z^{(1)}>0$; otherwise, it is negative. To compute $\phi$, we first select a node in the dimple (the blue sphere in Fig.~\ref{fig:sketch}) and then we compute the angle between the vector which connects the center of mass of the RBC to the node in the dimple $\vec{e}^{\tiny \mbox{D}}$ and the vector $\vec{e}^{(1)}$; the vector product $\vec{e}^{\tiny \mbox{D}}\times \vec{e}^{(1)}$ is used to determine whether $\phi$ is positive or negative (see Fig.~\ref{fig:sketch}).
In Fig.~\ref{fig:sketch-b} we plot the time evolution of the angles $\theta$ and $\phi$ for both TB (top) and TT (bottom): the TB is characterised by values of $\theta$ (red triangles) ranging in $[-90,90]$, while the angle $\phi$ (blue circles) oscillates; on the contrary, in the TT simulation, $\phi$ ranges in $[-90,90]$ and $\theta$ oscillates. The highest value of capillary number $\Ca$ that corresponds to a pure TB is identified by $\CaTB$, while $\CaTT$ represents the smallest value of $\Ca$ where a pure TT is found (see sketch in Fig.~\ref{fig:sketch_trans}). Values of $\Ca$ such that $\CaTB<\Ca<\CaTT$ identify the transition region (shaded region in Fig.~\ref{fig:sketch_trans}), where the dynamics starts as TB and switches to TT; the width of such region is defined by
\begin{equation}\label{eq:deltaca}
\Delta\Ca=\CaTT - \CaTB\; .
\end{equation}
In Fig.~\ref{fig:trans-a} and Fig.~\ref{fig:trans2}, we report only the error bars to identify the width of the transition region, otherwise there would be the superposition of shaded regions for different values of membrane viscosity $\mum$.

\section{Results}
To study the dependency of the TB-TT transition on the viscosity ratio $\lambda$ (see Eq.~\eqref{eq:visc_ratio}), the capillary number $\Ca$ (see Eq.~\eqref{eq:ca}) and the membrane viscosity $\mum$, we performed 3D numerical simulations to generate four different phase diagrams ($\lambda, \Ca$), each one related to a value of membrane viscosity $\mum$ in the range $[0,3.18]\times 10^{-7}\mbox{ m Pa s}$.
For each value of membrane viscosity $\mum$, and for each value of viscosity ratio $\lambda$, we consider the highest value of capillary number $\Ca$ which gives a pure TB motion ($\CaTB$) and the lowest value of $\Ca$ which corresponds to a pure TT motion ($\CaTT$): the difference $\Delta\Ca$ (see Eq.~\eqref{eq:deltaca}) represents the width of the transition region that is represented via error bars (see Sec.~\ref{sec:model}).\\
In Fig.~\ref{fig:trans1-a}, the phase diagrams $(\Ca,\lambda)$ at varying values of membrane viscosity $\mum$ is reported. First, we benchmarked our model against Cordasco~{\it et al.}~\cite{cordasco2014comparison} for $\mum=0$ (see black line in Fig.~\ref{fig:trans1-a}), and we found a good match. Even if the matching is good, it is not perfect: but as Cordasco~{\it et al.}~\cite{cordasco2014comparison} stated, the line they report to separate the two regions of TB and TT is approximate; moreover, the elastic modulus $\kS$ (see Eq.~\eqref{eq:skalak2}) we chose according to experimental data~\cite{art:suresh2005connections} is slightly different from theirs. Then, we found that increasing the membrane viscosity $\mum$, the transition line rises, i.e., for a fixed value of viscosity ratio $\lambda$, the TB-TT transition takes place for a higher value of the capillary number \Ca. 
In Fig.~\ref{fig:trans1-b}, we report such capillary number as a function of the membrane viscosity $\mum$, for different values of viscosity ratio $\lambda$. 
It follows that both membrane viscosity and viscosity ratio have the same {\it qualitative} effect on the TB-TT transition, that is to penalise the TT: the higher the value of membrane viscosity $\mum$ (and viscosity ratio $\lambda$), the higher the value of $\Ca$ such that the membrane tank-treads. In the limit of $\mum\to\infty$ (or $\lambda\to\infty$), i.e., in the limit of a rigid body, the membrane does not tank-tread. 
\\
It is also interesting to study the dependency of the width of the the transition region $\Delta\Ca$ as a function of membrane viscosity $\mum$ and viscosity ratio $\lambda$ (see Fig.~\ref{fig:trans2}). In Fig.~\ref{fig:trans2-a}, we report $\Delta\Ca(\lambda,\mum)$ as a function of $\lambda$ for fixed value of membrane viscosity $\mum$, finding an increasing function for all the different values of $\mum$. Moreover, the rate at which each curve increases, i.e., $\frac{\partial}{\partial\lambda}\Delta\Ca(\lambda,\mum)$, depends on the value of the membrane viscosity $\mum$. In Fig.~\ref{fig:trans2-b}, we report the width of the transition region $\Delta\Ca(\lambda,\mum)$ at varying viscosity ratio $\lambda$: similarly to the previous case, $\Delta\Ca(\lambda,\mum)$ increases with the membrane viscosity $\mum$, and the rate of increase $\frac{\partial}{\partial\mum}\Delta\Ca(\lambda,\mum)$ is proportional to the viscosity ratio $\lambda$. Overall, we can say that the qualitative effect of the viscosity (either via the viscosity ratio $\lambda$ or via the membrane viscosity $\mum$) is to increase the width of the transition region; nevertheless, we note that there is a quantitative difference between the effects of $\lambda$ and those of the membrane viscosity $\mum$ (i.e., the two panels of Fig.~\ref{fig:trans2}): looking at Fig.~\ref{fig:trans2-b}, we can see that $\Delta\Ca(\lambda,\mum\to 0)$ approaches very small values regardless of the value of viscosity ratio $\lambda$; on the other hand, in Fig.~\ref{fig:trans2-a}, the width of the transition region $\Delta\Ca$ does not seem to go to comparably small values when $\lambda\to 0$; rather, it depends on the value of membrane viscosity $\mum$: $\Delta\Ca(\lambda\to0,\mum) = \Delta\Ca_0(\mum)$. In particular, the smaller the membrane viscosity $\mum$, the smaller $\Delta\Ca_0(\mum)$. These considerations suggest that the use of an effective viscosity ratio $\lambda'$ to simulate the effect of membrane viscosity $\mum$ on the TB-TT transition can only be qualitatively correct; therefore, a direct implementation of $\mum$ is needed for a more quantitative and realistic analysis. \\

\section{Conclusion}
In this work, we performed 3D numerical simulations in the framework of the Immersed Boundary - Lattice Boltzmann method to investigate the dependency of the TB-TT transition on the membrane viscosity $\mum$ for a single RBC. We studied the transition region, i.e., the region in the phase space $(\Ca,\lambda)$ where the dynamics is neither pure TB nor pure TT: in particular, in this region we found an initially TB dynamics which switches to TT; the contrary, i.e., a TT that switches to a TB, has never been observed (in agreement to~\cite{kessler_finken_seifert_2008,bagchi2009dynamics}). In particular, Cordasco~{\it et al.}~\cite{cordasco2014comparison} highlighted that the intermittent dynamics in the transition region depends on the stress-free configuration of the membrane, and for the biconcave discocyte shape they did not observe intermittency (while it was observed for a stress-free state close to a sphere). \\
We found that the effect of the viscosity (regardless of whether it is fluid or membrane viscosity) is to shift the TB-TT transition line in the phase space $(\Ca,\lambda)$: in particular, the higher the value of the viscosity ratio $\lambda$ or the membrane viscosity $\mum$, the higher the transition line (see Fig.~\ref{fig:trans-a}). We also found that the width of the transition region $\Delta\Ca$ depends on both the viscosity ratio $\lambda$ and the membrane viscosity $\mum$. 
Again, increasing $\lambda$ or $\mum$ qualitatively leads to an increase of the width $\Delta\Ca$; at a more quantitative level, we found that  the two viscous effects are not equivalent. In particular, we found a different behaviour for small values of $\lambda$ and $\mum$: on one hand, for small values of $\mum$, the width $\Delta\Ca$ becomes very small independently of the value of the viscosity ratio $\lambda$; on the other hand, for small values of $\lambda$ , the width $\Delta\Ca$ is not small and shows a substantial dependency on $\mum$ (see Fig.~\ref{fig:trans2}). This suggests that the direct implementation of membrane viscosity is a required ingredient for a precise and realistic description of RBCs. \\
On a future perspective, it could be interesting to develop a reduced model to better understand the physics underlying our observations and explain the functional behaviours that we have unveiled via the numerical simulations.


\paragraph*{Acknowledgment:} the authors acknowledge Giannis Koutsou.
\paragraph*{Funding:} this project has received funding from the European Union Horizon 2020 research and innovation programme under the Marie Sk\l odowska-Curie grant agreement No 765048. We also acknowledge support from the project ``Detailed Simulation Of Red blood Cell Dynamics accounting for membRane viscoElastic propertieS'' (SorCeReS, CUP N. E84I19002470005) financed by the University of Rome ``Tor Vergata'' (``Beyond Borders 2019'' call).


\bibliographystyle{ieeetr}
\bibliography{rsc}

\end{document}